# MICROSTRUCTURAL EVOLUTION DURING AGEING OF Al-Cu-Li-x ALLOYS

**Vicente Araullo-Peters*[a,b], Baptiste Gault[c], Frederic de Geuser[d], Alexis Deschamps[d], Julie M. Cairney[a,b]**

[a] School of Aerospace Mechanical and Mechatronic Engineering, University of Sydney, Sydney, Australia

[b] Australian Centre for Microscopy and Microanalysis, University of Sydney, Sydney, Australia

[c] Department of Materials, University of Oxford, Parks Road, Oxford, OX1 3PH, UK

[d] SIMaP, Grenoble INP-UJF-CNRS, Saint-Martin-d'Heres, France

*Corresponding Author Email: vicente.araullopeters@sydney.edu.au (V Araullo-Peters)

## Abstract

In this study atom probe tomography was used to investigate the microstructure of AA2198 (Al-1.35Cu-3.55Li-0.29Mg-0.08Ag) over a range of ageing conditions to examine the evolution of phases in the alloy, in particular aiming to reveal the nucleation mechanism of the strengthening T1 phase, which has been under debate for decades. $T_1$ precursor phases were observed from early ageing, a significant number of which were connected to solute-enriched dislocations. Ag and Mg segregation to $T_1$ interfaces was convincingly observed when the plates were oriented perpendicular to the probing direction, which is the condition under which the spatial resolution of the atom probe data is highest.

## 1. Introduction

Aluminium lithium alloys are of high-interest for commercial, military and aerospace applications due to their excellent combination of low density, high stiffness and high strength [1-3]. For aluminium alloys, Li provides the largest reduction in density and greatest increase in stiffness per percentage weight of any known alloying element [4]. Every 1 at % of Li added to the alloy reduces the total density of the alloy by 3% and the Li solubility in aluminium is approximately 4.2 at% [5]. When other solutes are added, particularly Cu, the strengthening potential of Al-Li alloys becomes remarkable as a result of the sophisticated combination of phases that form during processing (Table 1 lists some of the phases commonly found in these alloys). In order to improve the properties of Al-Li-Cu alloys through the design of appropriate compositions and thermomechanical processing regimes, it is therefore necessary to understand the complex microstructure within these alloys and the sequence of events that lead from the as-quenched supersaturated solid solution to the end microstructure.

The AA2198 alloy, whose composition is given in Table 2, has been shown to have a desirable combination of tensile properties, formability and damage tolerance [6]. These Al-Li-Cu alloys can be produced with high yield strength, which is attributed to the precipitation of the $T_1$ phase ($Al_2LiCu$). This phase forms as thin plates lying on the {111} matrix planes. Its structure as a bulk phase has been determined by Van Smaalen and co-workers [7] and its structure as an precipitate embedded in aluminium has been more recently established [8, 9]. There are four reported nucleation sites for the phase: grain boundaries, zirconium dispersoids, octahedral voids/gp zones, and dislocations [10]. Cold work prior to ageing has been shown to increase the amount of intra-granular precipitation of $T_1$ [11-13], which has been attributed to the increase in dislocation density and therefore potential nucleation sites in the form of high strain fields around kinks and jogs [14-16]. This is consistent with a model of $T_1$ nucleation where the plates are nucleated on the dissociation of a dislocation on opposite sides of a jog or cross-slipped screw segment [14].

Mg and Ag are known to aid $T_1$ precipitation and hence to improve the hardness of the alloy [13, 16, 17]. While a combination of Mg and Ag provide the greatest increase in hardness, alloying with Ag alone provides little gain. The distribution of these elements in the structure of the $T_1$ plates, as well as their partitioning between precipitates and matrix, as measured by atom probe, is the subject of conflicting reports. Some studies have shown the presence of Ag at the interface between the $T_1$ with the matrix [18], while others have not observed such interfacial segregation [19]. In a recent study, Decreus and co-workers [20] investigated in detail the precipitation sequence during ageing at 155°C in AA2198 alloy using Transmission Electron Microscopy (TEM) and Small-Angle X-ray Scattering 5SAXS). They provided evidence that the peak hardness microstructure consisted of thin $T_1$ platelets associated with a smaller fraction of Cu-rich phases (GPI, GPII and θ'). The sequence of events leading to this microstructure, as evidence by in-situ SAXS observations, involved first the dissolution of low-temperature clusters during the heating ramp to the ageing temperature, followed by an incubation time where no significant precipitation could be detected by this technique and, after about 2h at 155°C, sudden apparition and growth of the $T_1$ precipitates. From the available data, there was no indication about the mechanisms that could precede and accompany the nucleation of $T_1$ during this incubation time and during the subsequent precipitate growth.

The aim of the present study is to use state-of-the-art atom probe tomography and innovative data treatment methods to provide new insights into the micro-structural evolution of the AA2198 structure at different ageing times, with particular emphasis on the early stages of precipitate nucleation when global characterisation techniques such as SAXS [Decreus et al] do not provide evidence of any detectable precipitation. Atom probe tomography is capable of determining the location and mass-to-charge ratio of atoms within volumes of approximately 100 nm x 100 nm x 500 nm with sub-nanometer resolution and can also be used to determine the orientation of the grains and the crystallographic features within them [21-26]. It is therefore a useful tool for examining the orientation and composition of the micro-structural features in the early stages of ageing in the AA2198 alloy. The key concerns of this study are the location of Mg and Ag within the $T_1$ structure, the various $T_1$ nucleation sites and the interaction between $T_1$ precipitates, dislocations and other phases throughout all ageing conditions.

## 2. Experimental Preparation and Methods

Samples of AA2198, an alloy in the AIRWARE® family with a composition given in Table 1, were provided by the Constellium-Voreppe Research Centre, France. The AA2198 alloy was received in the T351 temper before being further aged. This involves solution treatment and water quenching, tensile predeformation to about 2% strain and then natural ageing for several months. The ageing treatment included a 20 K h$^{-1}$ ramp up to 155°C followed by isothermal annealing for 0, 1, 2, 15 or 84 hrs.

Atom probe specimens were prepared by standard eletropolishing techniques with 25% perchloric acid and 70% glacial acetic acid at 14V. Atom probe experiments were conducted with a Cameca LEAP® 3000X Si instrument operating in high-voltage pulsing mode at a base temperature of 20 ±5K and with a HV pulse amplitude 20% of the DC voltage applied to the specimen. Data reconstruction, using standard reconstruction algorithms [27], and 3D visualisation was carried out using the commercial software IVAS$^{TM}$ (Cameca). Calibration of the tomograms was carried out using the methods introduced in refs [28, 29].

The methods used for crystallographic characterisation of specimens are described in detail in [21, 25, 30]. They make use of sets of atomic planes present within the data that are often associated with so-called 'poles' that take the form of lower- or higher-density regions in the data as shown in Figure 2(a). After identifying each pole in the atom probe data, the orientation of the crystallographic planes associated with each pole is found with respect to the analysis direction by applying a 3D Hough Transform algorithm to the dataset, as outlined in [25]. This provides sufficient information to find the orientation of crystal grains in the atom probe data with respect to the analysis direction. The orientation of features with respect to the grain itself is then determined.

## 3. Results and discussion

*3.1 End of Ramp*

At the end of the heating ramp, no significant clustering or precipitation can be detected by averaging methods such as SAXS or differential scanning calorimetry [20]. However, atom probe tomography reveals that the microstructure is not completely homogeneous. The microstructure consists of dislocations enriched with Cu and Mg (Figure 1 and Figure 2) and a low density of small platelets that contain Cu, Li, Mg and Ag (Figure 2). While most datasets at this ageing condition do contain either a dislocation and/or platelet feature (5 out of 8 datasets collected at this ageing condition), these features are sparsely distributed throughout the data. Dislocations are identifiable in the atom probe images as Cu/Mg-rich curved lines or 'strands', winding through the datasets and often forming loops. Analysis of the crystallographic orientation of the dataset allows the plane on which the dislocations are lying to be determined. Figure 1(a) is a 2D density map normal to the analysis direction. The apparent low density areas are traces of crystallographic directions ("poles") from which the orientation of the crystal grains in the dataset can be determined. Figure 1(b) shows an example of a dislocation loop that intersects the dataset and it has been determined that it is lying on a (110) plane. The volume shown is ~22 nm tall, 25 nm long and around 8 nm thick and is a small section of a larger atom probe dataset. The faces of the volume shown correspond to the {110} planes within the sample from which the data is obtained. The solute-rich region around the dislocation has been highlighted with the use of isoconcentration surfaces of Cu (orange) and Mg (green).

At this heat treatment, dislocations loops were observed on both the {110} and {111} planes. Those loops forming on {111} planes are consistent with dislocations that are the result of deformation and those loops lying on the {110} planes are consistent with quenched-in dislocation loops [31].

Figure 2 is a characteristic area that shows both a curved Cu/Mg-rich line-like feature and 2 small platelets. Figure 2(a) and Figure 2(b) are the same volume seen from two different angles. The platelets are typically between 5 and 10 nm diameter and crystallographic analysis of the data revealed that they lie on {111} planes. Figure 2(c) is a concentration profile through one such platelet, which is found to be enriched with Cu, Li, Mg and Ag (in order of decreasing at. %). On the basis of the composition and orientation, these platelets are considered to be early-stage $T_1$ precipitates. Figure 2(d) is a proximity histogram [32] (proxigram) from the isoconcentration surface of the dislocation, which shows the average composition moving away from the Cu isoconcentration surface depicted in green, and labelled 'CuMg' in Figure 2(a) and Figure 2(b). Because the region around the dislocation is rich in Cu and Mg, with only a small enrichment in Li and Ag in these areas, the composition is thought to be most similar to that of the S phase or some of its precursors (GPB zones).

In most cases investigated, the concentration of solute elements along the dislocations lines is not uniform. Qualitatively, segments can be seen to have higher Li, Cu and Ag concentrations, especially in the regions in which the dislocations are most curved. Figure 3 shows a set of 2D concentration profiles over the same region in an atom probe data set from a specimen aged to the end of ramp. This particular dataset contains a line feature (a dislocation) which is comprised of a relatively straight section and a curved section. The concentrations distributions for Al, Li, Cu, Mg and Ag are shown. Depletion of Al and segregation of Cu and Mg is observed over the whole dislocation. At the curved section of the dislocation there is an increased level of segregation of Cu and Mg as well as some Li and Ag.

### *3.2 1hr and 2hr ageing*

From macroscopic observations 1 and 2 hours of ageing are thought to be during the incubation time for $T_1$ precipitation and at the beginning of generalized $T_1$ precipitation, respectively [20]. Figure 4 shows a dataset that contains typical structures seen by APT after ageing in these two conditions. The atom probe data set shown in this figure is from a sample annealed for 1 hour at 155°C. The figure shows four different views of the same region of interest, which contains a line feature, believed to be a dislocation loop and a (Cu,Li)-rich plate, which can be identified as a $T_1$ precipitate and lies on the {111} matrix plane. In Figure 4(a), (c) and (d), the Mg and Cu isoconcentration surfaces are respectively green and orange. In Figure 4(b) a Ag isoconcentration surface is represented in black. It highlights a localized Ag enrichment at the dislocation which may be an early indication of $T_1$ nucleation. The $T_1$ plate labelled on the right-hand side of the image is approximately 15 nm long and is attached to a Cu/Mg-rich dislocation at each end. Unlike the dislocation loop on the left, this dislocation is not lying on the same plane as the precipitate.

For the specimens aged for one and two hours, the microstructure is dominated by dislocations with plates attached such as those shown. The plates attached to the dislocations were consistently found to be lying on {111} planes and have a composition close to that expected for $T_1$ plates. In some cases, these dislocations (with plates attached) are aligned, presumably forming a sub-grain boundary. Compared to the end of ramp heat treatment, the dislocations after one and two hours are also Cu- and Mg-rich but have significantly more segregation. The segregated regions occasionally appear to protrude along the <100> direction. This is best seen for the loop on the left in figure 4(c). This would be consistent with the start of the formation of S laths, which have a {102} habit plane and extend in the <100> directions and have been previously observed to nucleate at dislocations [33].

Large Zr-rich dispersoids were also observed in the data for the specimens annealed for one and two hours, which are identified as β' phase. In several cases, $T_1$ plates were attached to these precipitates. Figure 5 shows one such precipitate with two small $T_1$ plates attached to it. The β' precipitate is marked by a purple Zr isoconcentration surface and the $T_1$ plates are marked by orange Cu isoconcentration surfaces. These β' precipitates are known to contribute only minimally to the strengthening of the alloy, but instead contribute to grain refinement [34].

θ' plates or their precursors were not observed at ageing times of two hours or less.

### 3.3 15hr and 84hr ageing

According to macroscopic observations, 15 and 84 hours of ageing correspond respectively to the beginning and the middle of the peak strength plateau at 155°C [20]. Little difference was observed between the type and size of the structures observed by APT in the specimens aged for 15 and 84 hours.

Figure 6(a), (b) and (c) is an area of a data set from a specimen aged for 84 hours, showing three different views of the same region. It contains two (Cu,Li)-rich plates on the {111} matrix planes that have been identified as $T_1$ and are highlighted by the orange Cu isoconcentration surface. It also contains two Cu-rich plates lying on the {100} matrix planes that have been identified as $\theta'$ plates and are indicated by the pink Li isoconcentration surface. Finally, it contains two (Cu,Mg)-rich laths that can be identified from their composition and orientation as S precipitates (or the precursor S') and are shown by the green Mg isoconcentration surface. Crystallographic orientations for two of the views (figure 6 (a) and (c)) are also shown.

Figure 6 also shows a dislocation loop with S-phase laths growing off it in the [100] direction. The loop is labelled in Figure 6(c), and the S-phase can be most clearly seen in Figure 6(a) and (b). The two $T_1$ plates, both lying on the {111} matrix plane, as well as two $\theta'$ plates, which are lying on two different {100} planes, are all connected to this loop, which is likely to be a common nucleation site.

$\theta'$ plates were observed in the materials aged 15 and 84 hours. Figure 7 shows a section of an a data set containing a GP zone, $\theta'$ plate and a $T_1$ plate. Cu atoms are shown in red and a 10% portion of the Al atoms in the dataset are shown in purple. The $\theta'$ plates were enriched in Ag, similar to the $T_1$ phase, but contained no enrichment of Mg. The GP zones showed no Mg or Ag enrichment. They also have a coating of $\delta'$ which has been reported before in [19, 20, 35-38]. This is the only form in which $\delta'$ was observed in this study with the exception of clumping to $\beta'$ dispersoids. Overall, a much lower number of $\theta'$ plates were observed compared to $T_1$ plates. To provide some indication of the frequency with which each phase was observed, in the 45 datasets acquired, 8 $\theta'$ plates were observed compared to 98 $T_1$ plates. Also, the fact that these plates were never observed in end of ramp, 1 or 2 hour ageing times is significant. They either cannot form due to competition with the growing $T_1$ phase, or they take longer to nucleate. This is, of course, especially surprising for GP zones, which normally appear very early in the Al-Cu precipitation sequence. This fact provides evidence that even after 84h ageing, a significant supersaturation of the matrix remains.

The S-related phases are seen in a variety of different forms. Typically S phase laths lay on {210} planes and these were regularly seen to intersect with $T_1$ plates (see Figure 11, discussed later). In Figure 6, attached to the edge of the upper θ' plate, is a S lath which appears corrugated, seemingly indicating that it consists of a combination of rods which are all parallel with a <100> direction. This configuration of S and θ' sharing a common edge was observed in multiple data sets. Also in Figure 6, S phase is seen growing off the dislocation loop as rods and it appears that the $T_1$ plate has blocked the growth of some of these rods.

Like for the earlier ageing conditions, β' dispersoids were observed at 15 and 84 hours. They were characterised by Li, Cu and Mg segregation as clumps attached to the β' dispersoids.

### *3.4 Mg and Ag segregation in $T_1$ plates*

Mg and Ag enrichments were observed in all $T_1$ precipitates in this study. As mentioned in the introduction, the precise location of Ag and Mg in these $T_1$ plates has been a point of contention in various studies. Murayama *et al* [18] observed segregation of Ag and Mg to the surface of these $T_1$ plates but this was not observed in the atom probe study of Gault *et al.* [35]. Figure 8(a), (b) and(c) show atom probe data cross-sections though $T_1$ plates which were observed in close proximity to <111> poles and parallel to the {111} matrix planes. The Al, Ag and Mg atoms are shown in blue, black and green respectively. Figure 8 (a), (b) and (c) correspond to datasets from specimens at the end of ramp, 2 hr and 84 hr ageing conditions. It can be seen that there is a clear segregation of Ag and Mg to the surface of these $T_1$ plates. In figure 9c, the centre of the image corresponds to a {111} crystallographic pole in the atom probe data set, a region where the probing direction of the atom probe is parallel to the <111> crystallographic direction and the plate is perfectly perpendicular to the probing direction. As one moves away from the pole one observes the segregation of the Ag and the Mg appears weaker as the plate is no longer perpendicular to the probing direction of the atom probe. This is due to the evaporation sequence, and hence the determination of the depth coordinate, becoming less well determined the further away from major atomic terraces [39]. This segregation, though seen at all ageing times, was only observed when plates were found in the vicinity of the {111} poles, as these are the habit planes of the $T_1$.

### *3.5 Discussion : nucleation sequence leading to $T_1$ precipitates*

Results such as the data shown in figure 5 provide strong indication of $T_1$ plates nucleating on parts of dislocations. In addition, highly curved regions of dislocations were observed to have higher Cu, Ag and Li concentration (Figure 4). The curved Mg- and Cu-enriched tubes have very similar composition to the S phase ($Al_2CuMg$) and it is thought that these are likely nucleation sites for the formation of this phase.

Close examination of the regions in which the $T_1$ plates intersect S phases provides evidence that further supports the theory of $T_1$ nucleation on dislocations and, in this case, specifically on jogs and kinks. In Figure 9, the dislocation, shown by a green 3.2% Mg isoconcentration surface, appears to be kinked at the point where there is a $T_1$ plate, shown by the orange 12% Cu isoconcentration surface. It is believed that this $T_1$ precipitate has nucleated at the location of the kink, leading to this resulting structure. It is therefore supposed that, at early ageing times, Mg and Cu solute atoms segregate to dislocations. At sharp points, kinks and jogs, there is more segregation of Cu and Mg, along with Li and Ag segregation. At 1 and 2 hours, plates grow from these areas along {111} planes and this growth continues to the 15 and 84 hr ageing times.

However, in some datasets from the specimens at early ageing times (0, 1 and 2 hours), plates were occasionally observed that did not have obvious nucleation sites. These were observed in 3 separate data sets (out of 26 collected), even in end of ramp specimens. They were seen to lie on {111} planes, have a $T_1$ composition and also display Mg and Ag segregation to the interfaces. These plates may have nucleated homogeneously, or from defects undetectable by atom probe, such as dislocations that have no visible segregation.

$T_1$ plates were also observed growing off $Al_3Zr$ precipitates, consistent with the observations from ref. [10]. $T_1$ plates were also observed attached to subgrain boundaries at 2 hours 84 hours which confirm that low angle grain boundaries are effective nucleation sites, as seen in previous TEM studies [10].

### 3.6 Intersection of Precipitates

Intersecting $T_1$ plates were also observed. Due to the fully three-dimensional nature of atom probe data, it was possible to unambiguously observe these $T_1$-$T_1$ intersections in 9 different datasets. Figure 10 shows one such intersection, which extends for a significant length. The $T_1$ plates are marked by an orange Cu 4% isoconcentration surface. $T_1$-S intersections were also commonly observed. Figure 11 is an atom probe dataset where an S lath (made visible by a green Mg 2.6% isoconcentration surface) cuts completely though one $T_1$ plate and impinges upon another (shown by a Cu 6.9% isoconcentration surface). This is consistent with the intersection of plates observed in previously [14, 35] and has been observed with electron tomography in [40].

### 3.7 Solute Composition in the matrix

Figure 12 displays the concentrations of the Li, Cu, Mg and Ag solute atoms in the matrix at each time, averaged over a large range of collected datasets. These concentrations were calculated from nearest neighbour distributions using the methods outlined in [41]. The concentration of the solute atoms in the matrix decreases to a stable level between the end of ramp and the 84 hour ageing times, including Ag and Mg. This decrease is attributed to the solute being incorporated into the $T_1$, S and θ' phases. In [35], a similar observation was made, however for long ageing times Mg and Ag solutes were observed to revert into solid solution, which was attributed to a catalytic effect of these species on $T_1$ precipitation. This was not observed in the present study, however their ageing temperature was 200°C, where precipitation is much faster than at 155°C (as used here) where the precipitation kinetics is very slow, especially when considering precipitate thickening as noted by [20]. It is therefore possible that at 155°C reversion of the minor solutes would also happen for very long ageing times where significant precipitate thickening occurs. The association of this catalytic effect with precipitate thickening would indicate that Mg and Ag not only serve as nucleation catalyst, but to a certain extent are also associated with the coarsening (i.e. thickening) resistance of $T_1$.

## 4. Conclusions

In this study we have used atom probe tomography to investigate the microstructure of the AA2198 alloy across a range of ageing conditions. The main findings are as follows:

- In early ageing conditions, Mg and Cu co-segregation is observed to occur at linear features that we identify as dislocations.
- Even in very early ageing conditions $T_1$ plates are observed, most of which are attached to dislocations and sub-grain boundaries, but some have no visible nucleation site. These plates often occur on parts of dislocations that are curved or kinked and in association with co-segregation of Cu and Mg, and to a lesser degree Ag. At later ageing stages, intersections of dislocation loops with $T_1$ plates, provide further evidence for the dislocation/$T_1$ plate interaction.
- In later ageing stages, the co-segregation of Cu and Mg at the dislocations is observed to develop into S precipitates as identified from their orientation with respect to the matrix. Intersecting $T_1$ plates are observed along with intersections between $T_1$ plates and S precipitates.

- Mg and Ag segregate to the in-plane matrix-$T_1$ interface at all ageing times, including the earliest stages. It is demonstrated that when the $T_1$ plate is not perpendicular to the atom probe probing direction the segregation to the interface becomes less clear, which explains previous reports of there being no observable segregation to the interface.

- Large, thin GP zones (or $\Theta''$) and $\Theta'$ plates were observed at later ageing times, but not at early ageing times which could indicate that these plates nucleate later, due to remaining Cu supersaturation.

- The evolution of Cu, Li, Mg and Ag in solid solution follow the same kinetics, which is consistent with the fact that these solutes are incorporated together in the most prevalent precipitate, namely $T_1$. Cu depletion in the solid solution surrounding the precipitates is observed in all ageing conditions investigated, indicating remaining Cu supersaturation.

This set of observations evidences the complex interaction of solutes in this multi-constituent alloy, and particularly shows the important role played by the minor solutes (Mg, Ag) on the nucleation of the $T_1$ phase.

## Tables

| Phase | Composition | Crystal Structure | Common Morphology |
|---|---|---|---|
| T1 | $Al_2CuLi$ | Hexagonal | Plates |
| $\Theta'$ | $Al_2Cu$ | Tetragonal | Plates |
| $\delta'$ | $Al_3Li$ | $Ll_2$ | Coating |
| $\beta'$ | $Al_3Zr$ | $Ll_2$ | Dispersoid |
| S | $Al_2CuMg$ | Orthorhombic | Laths, rods |

**Table 1 Common phases in the Al-Li-Cu-Mg-Ag-Zr alloys and the Composition of the AA2198 alloy**

| Element | Li | Cu | Mg | Ag | Zr |
|---------|-----------|-----------|------------|-----------|-------------|
| Wt.%    | 0.8 - 1.1 | 2.9 - 3.5 | 0.25 - 0.8 | 0.1 - 0.5 | 0.04 - 0.18 |

Table2

# FIGURES

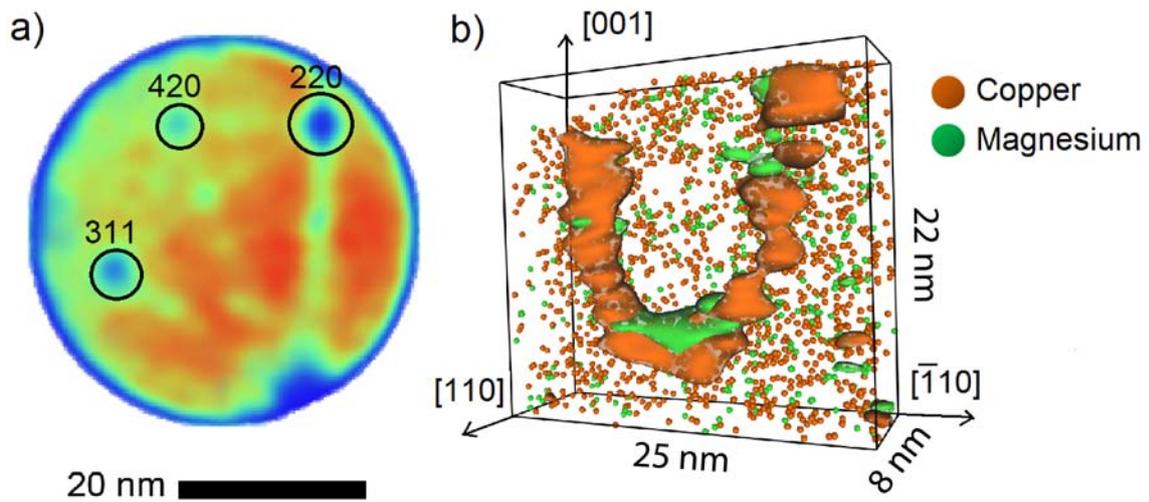

Figure 1 (a) A 2D density map of Al from a slice of the atom probe data set, depicting the poles and therefore the orientation of the crystal grain. (b) A 3D map showing part of a dislocation loop in an as-received AA2198 sample. A fraction of the Cu (orange) and Mg (green) atoms are shown, together with isoconcentration surfaces (Cu 3.6% in orange and Mg 1.6% in green) that highlight the loop due to the segregation of these elements.

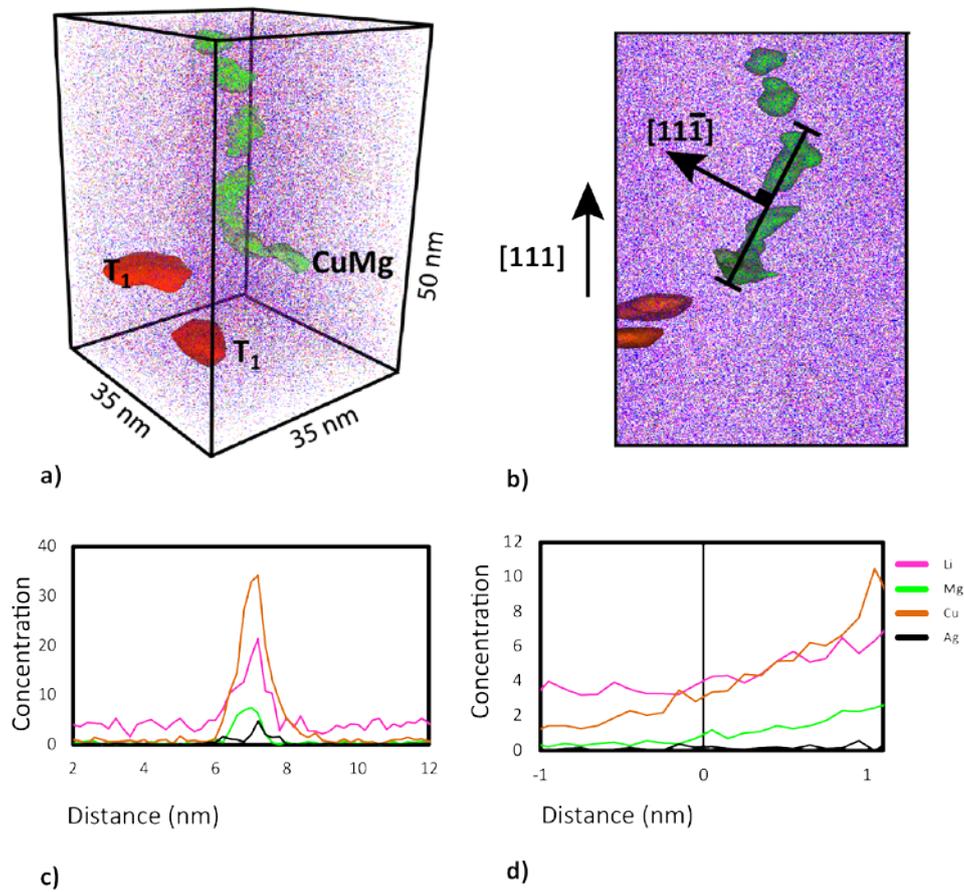

**Figure 2 (a),(b): Atom probe data of an end of ramp specimen containing $T_1$ precipitation in orange and a MgCu rich tube, believed to be a dislocation, in green. The purple dots are Al atoms (only a small fraction are shown) (c) A concentration profile though one of the T1 precursors, (d) A proxigram showing an increase in Li Mg Ag and Cu near the dislocation. Both the green and orange are 4% Cu isoconcentration surfaces coloured differently to distinguish between them.**

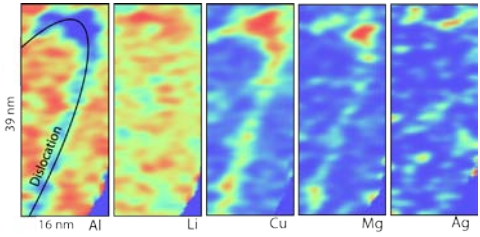

Figure 3 2D concentration profiles of an end of ramp atom probe data set. Al, Li, Cu, Mg, Ag shown for a dislocation with a curved section.

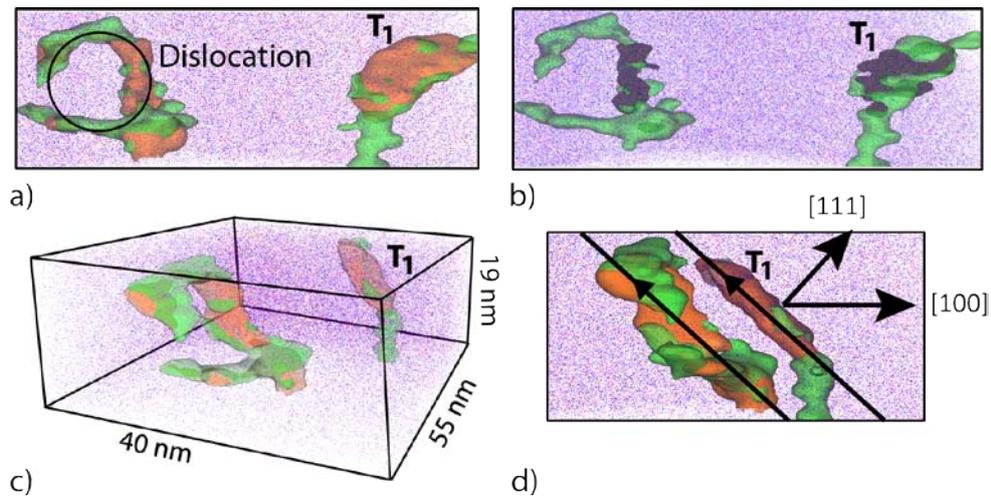

Figure 4 Atom probe data from a specimen aged for 1hr. Green shows Mg-rich regions (4.16 Mg isoconcentration surface), believed to be dislocations, and orange shows Cu-rich regions (1.27 Cu isoconcentration surface), believed to be places where $T_1$ phase is growing. Black regions are Ag-rich (Ag, 0.7% isoconcentration surface).

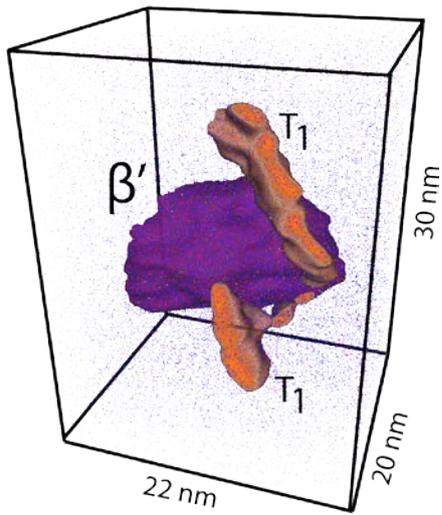

**Figure 5** 1hr Atom probe data containing a β′ precipitate with two $T_1$ precipitates attached. The purple isoconcentration surface is Zr 1.6% and the orange isoconcentration surface is 5.5% Cu+Mg.

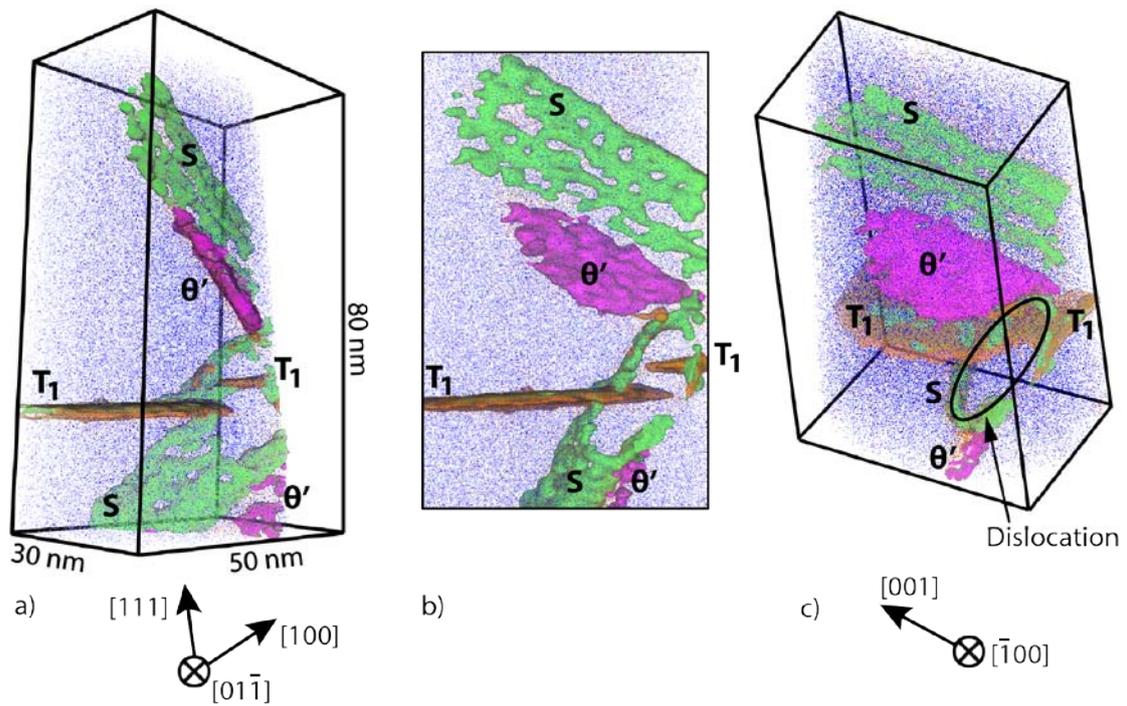

**Figure 6** Atom probe dataset from the specimen aged for 84 hours. The different objects were identified by their composition and coloured according to their phase. The S phase is shown in green, θ' phase shown in pink and $T_1$ plates shown in orange. All the plates are connected to the dislocation loop, which is highlighted by the S phase which that is growing from it.

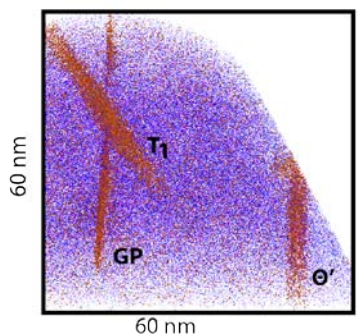

**Figure 7** Atom probe data of 84 hr specimen showing a GP zone, $T_1$ plate and θ' plate. Cu atoms are shown in red and Al atoms are shown in blue. Only 10% of the Al atoms were shown. The GP and θ' plates observed were parallel.

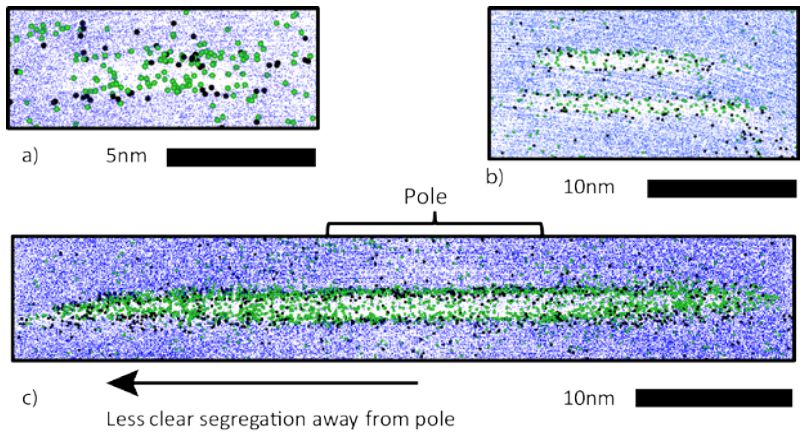

**Figure 8** Atom probe data cross sections of various $T_1$ plates found to lie on 111 planes at (a) end of ramp,(b) 2hrs and (c) 84hrs. Al-blue, Mg-green, Ag-Black. The different ageing conditions show clear segregation at the pole, which deteriorates as one moves away from being parallel with the probing direction

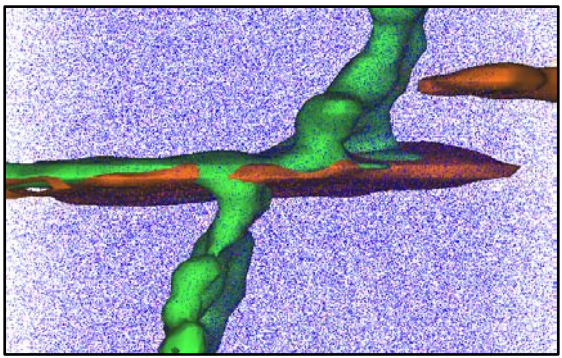

**Figure 9** A connection between the dislocation loop and a $T_1$ plate. Green is a Mg 3.3% isoconcentration surface and orange is a Cu 12.0% isoconcentration surface. The dislocation, depicted in green, is kinked at the point where it meets the T1, orange.

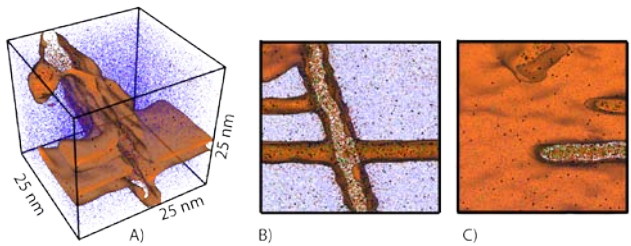

**Figure 10** Atom probe data of an 84 hour aged specimen containing intersecting $T_1$ plates. (a) Perspective view, (b) & (c) orthogonal views demonstrating an intersection of $T_1$ plates. The orange is a Cu 4% Isoconcentration surface.

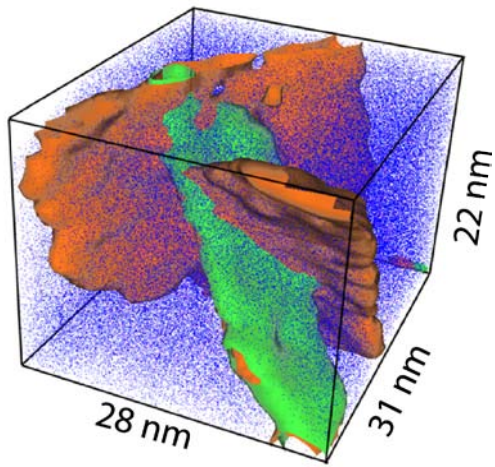

**Figure 11** Atom probe data of an 84 hour aged specimen. Region containing a S lath shown with a green 2.4%Mg Isoconcentration surface intersecting with two $T_1$ plates shown with orange 6.9% Cu isoconcentration surfaces.

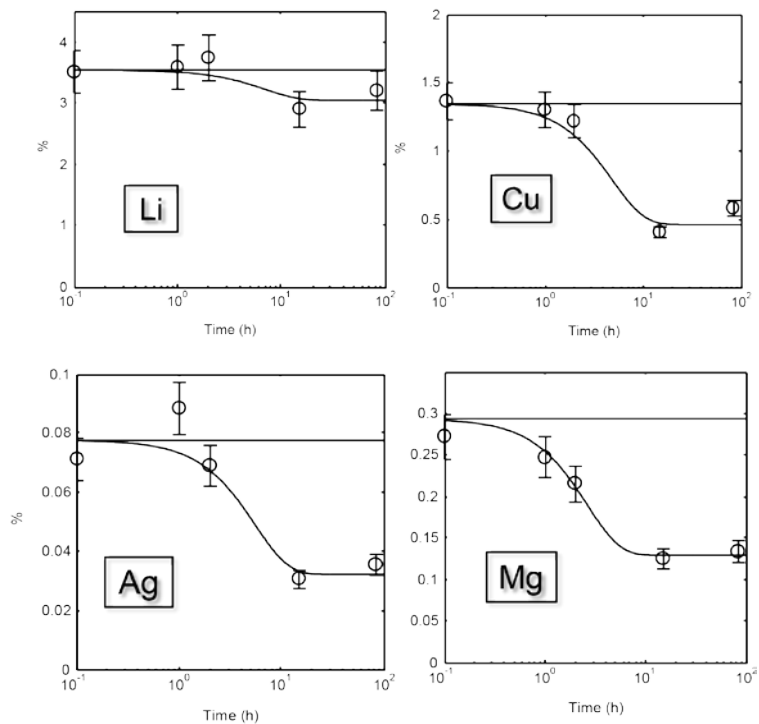

**Figure 12** Average matrix compositions for each ageing time. The black lines are simply provided as a guide for the eye.